\documentclass[a0paper, 11pt, oneside]{article}
\usepackage[a4paper, total={6.5in, 9in}]{geometry}

\usepackage{graphicx}
\usepackage[np]{numprint}
\npthousandsep{\,}
\npdecimalsign{.}
\npproductsign{\times}
\npfourdigitnosep
\usepackage{physics}
\usepackage{amsfonts}
\usepackage{amsmath}
\usepackage{amssymb}
\usepackage{slashed}
\usepackage{booktabs}
\usepackage{color}
\usepackage[hidelinks]{hyperref}
\usepackage[font=footnotesize, labelfont=bf]{caption}
\usepackage{indentfirst}

\definecolor{charm}{RGB}{255,0,0}
\definecolor{ps191}{RGB}{0,145,45}
\definecolor{peak}{RGB}{220,150,40}

\newcommand{\tapi}{\textsuperscript}

\newcommand{\ra}{\rightarrow}

\newcommand{\pd}{\partial}

\newcommand{\sh}{\slashed}

\newcommand{\cj}{\overline}

\newcommand\pubnumber{NuPhys2017-Boschi}
\newcommand\pubdate{\today}

\let\defaultfootnote=\thefootnote
\renewcommand{\thefootnote}{\fnsymbol{footnote}}

\def\ippp{Institute for Particle Physics Phenomenology, Department of Physics,\\
	Durham University, South Road, Durham DH1 3LE, United Kingdom.}
\def\pprc{Particle Physics Research Centre, School of Physics and Astronomy, \\
	Queen Mary University of London, Mile End Road, London E1 4NS, United Kingdom.}
\def\support{\footnote[2]{This work is supported by the European Research Council under %
		ERC Grant ``NuMass'' (FP7-IDEAS-ERC ERC-CG 617143).}}

\def\Title#1{\begin{center} {\Large #1 } \end{center}}
\def\Author#1{\begin{center}{ \sc #1} \end{center}}
\def\Address#1{\begin{center}{ \em #1} \end{center}}

\newcommand\pubblock{\rightline{\begin{tabular}{l} \pubnumber\\
         \pubdate  \end{tabular}}}
\newenvironment{Abstract}{\begin{quotation}  }{\end{quotation}}
\newenvironment{Presented}{\begin{quotation} \begin{center} 
             PRESENTED AT\end{center}\bigskip 
      \begin{center}\begin{large}}{\end{large}\end{center} \end{quotation}}

\def\beq{\begin{equation}}
\def\eeq#1{\label{#1}\end{equation}}
\def\eeqn{\end{equation}}

\def\beqa{\begin{eqnarray}}
\def\eeqa#1{\label{#1}\end{eqnarray}}
\def\eeqan{\end{eqnarray}}

\let\bar=\overbar

\def\Dslash{\not{\hbox{\kern-4pt $D$}}}
\def\dslash{\not{\hbox{\kern-2pt $\del$}}}

\def\msb{{\bar{\ssstyle M \kern -1pt S}}}

\begin{document}
\begin{titlepage}
\pubblock

\vfill
\Title{Searching for MeV-scale Neutrinos with the DUNE Near Detector\support}
\vfill
\Author{Peter Ballett\tapi{a}, Tommaso Boschi\tapi{a,b}, Silvia Pascoli\tapi{a}}
\Address{\emph{A}. \ippp\\\emph{B}. \pprc}
\vfill
\begin{Abstract}
	Adding right-handed neutrinos to the Standard Model is a natural and simple extension %
        and is well motivated on both the theoretical and the experimental side.
	We extend the Standard Model by adding only one right-handed Majorana neutrino and study %
	the sensitivity of the Near Detector of the DUNE experiment to the new physics parameters, namely the mixing parameters %
        $|U_{e 4}|^2$ and $|U_{\mu 4}|^2$ and the mass $m_N$.
	The study relies on searches of the products of right-handed neutrino decays, which is possible thanks to %
	an extremely intense beam and a state-of-the-art detection technology.
	This type of direct test is carried out with very few assumptions and in an almost-completely model-independent way, %
	providing thus a strong result.
	A background analysis is also performed, simulating the detector performance to particle identification.
	It is found that the existing bounds in the MeV-range can be improved by one order of magnitude in different %
	detection channels.
\end{Abstract}
\vfill
\begin{Presented}
	NuPhys2017, Prospects in Neutrino Physics
	Barbican Centre, London, UK,  December 21, 2017
\end{Presented}
\vfill
\end{titlepage}
\def\thefootnote{\fnsymbol{footnote}}
\setcounter{footnote}{0}

\let\thefootnote=\defaultfootnote
\section{Introduction}

	The evidence for three neutrino flavour oscillation is well established~\cite{SuperK, SNO} %
	and can be accounted for only if the neutrino mass splittings are non zero~\cite{NuFit18}.
	This imply that at least two neutrino states are massive, %
	even though the Lagrangian of the Standard Model (SM) does not give an explanation of such masses.
	A natural solution is to extend the SM with the right-handed counterpart of neutrinos, in order to %
	accommodate in the Lagrangian both Yukawa couplings, which generate Dirac masses via Higgs mechanism, %
	and Majorana mass terms.
	A seesaw realisation can then take place to address the small neutrino masses, but its discussion is %
	beyond the scope of this work.

	The Deep Underground Neutrino Experiment (DUNE) will also be sensitive to Beyond the Standard Model (BSM) physics.
	In this study, we will show that it has the capability of performing a model independent search %
	for right-handed neutrino decays into SM particles.
	Thanks to very few assumptions, this type of analysis gives very strong limit on new physics content %
	in case of non-observation of such decays.
	In laboratory searches, no positive evidence of sterile neutrinos has been found so far in the mass range of interest.
	The current limit in this region is given by the \mbox{PS191} experiment~\cite{Bernardi88}.
	The upper bound was set to be $\leq \np{e-8} \sim\np{e-9}$ for both $|U_{e4}|^2$ and $|U_{\mu4}|^2$.
	A thorough review of the current constraints is found in~\cite{Atre09, Drewes17}.
	
% Head 2
\section{Theoretical framework}

	The addition of a single sterile fermionic state $N'$ to the SM is an economical modification and sufficient to describe %
	the search of neutrino decays into SM particles.
	Yukawa couplings between the new fermions and the Higgs field are permitted, as well as Majorana mass terms.
	The Lagrangian reads:
	\begin{equation}
		\label{eq:model}
		\mathcal{L} = \mathcal{L}_\text{SM} + i \cj{N'} \sh{\pd} N' - \sum_\alpha Y_\alpha \cj{L}_\alpha \tilde{H} N' - %
		\frac{1}{2} M_R \cj{N'^C} N' + h.c.\ ,
	\end{equation}
	with $\tilde{H} = i \sigma_2 H$ and $N'^C$ the charge conjugated field of $N'$.
	After the electroweak symmetry breaking and the diagonalisation of the mass matrix, %
	the new Majorana state $N$ mixes with the light ones into active neutrinos as:
	\begin{equation}
		\label{eq:massmix}
		\nu_\alpha = \sum_{i = 1}^3 U^*_{\alpha\,i}\ \nu_i + U^*_{\alpha\,4}\ N^C\ ,
	\end{equation}
	where the entries $U_{\alpha\,i}$ corresponds to the PMNS matrix and $U_{\alpha 4}$ are the new mixing elements.
	As the flavour eigenstates are coupled to the electroweak bosons, the new mass eigenstate is involved in any process in which %
	active neutrinos take part, with vertices proportional to the mixing $U_{\alpha 4}$.
	The masses of the new fermions can range from below the electroweak scale up to the Planck scale, %
	producing accordingly a different phenomenology.

	\begin{figure}
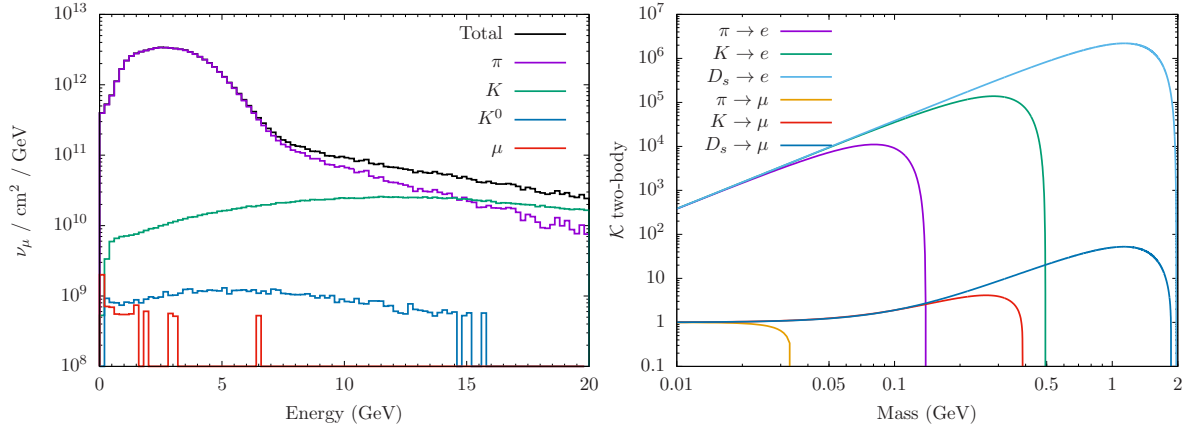

		\resizebox{0.5\textwidth}{!}{\input{fluxnumu.tex}}
		\hspace{-2em}
    	 	\resizebox{0.5\linewidth}{!}{\input{Two.tex}}
   		\caption{Left: prediction of the $\nu_\mu$ flux at the Near Detector. The spectrum is broken down into %
			the contribution coming from different parent particles.
			Right: factors $\mathcal{K}$ for two body decays of pseudomeson, up to the $D_s$ meson, plotted as a function of the mass.
			The enhancement in the electronic channel is due to the fact that %
			the Majorana neutrinos are not affected by helicity suppression.}
		\label{fig:branch}
	\end{figure}

	If the mass is the range from few MeV to few GeV, the heavy neutrinos can be produced in the beam, %
	in a fixed target experiment, with the same mechanism that generates light neutrinos.
	A proton beam impinging on a fixed target produces a significant number of light pseudo-scalar mesons, %
	mainly pions and kaons, which in turn decay via leptonic and semi-leptonic channels.
	The prediction of the $\nu_\mu$ spectrum broken down by meson parentage is illustrated in figure Fig.~\ref{fig:branch} on the left.
	The deflection of the mesons with magnetic horns into a decay pipe results in a focused neutrino beam, a component of which %
	consists of sterile neutrinos with masses $m_N$ up to the mass of the decaying meson.
	If the new neutrinos are massive enough, their mass-splittings with the light neutrinos could be larger than the wave packet %
	energy-uncertainty associated with the production mechanism, and so they do not oscillate~\cite{Akhmedov09}.
	In this scenario, MeV-scale sterile neutrinos can propagate undisturbed from the decay pipe to the detector, %
	where they might decay into visible SM particles.
	For masses $m_N < m_{K^0}$, the available decay modes are, in ascending mass threshold order, 
	\[
		N\ \longrightarrow\ 3\nu\ ,\ \nu\gamma\ ,\ \nu e^+ e^-\ ,\ \nu e^\pm \mu^\mp\ ,\ \nu \pi^0\ ,\ e^- \pi^+\ ,\ \nu \mu^+ \mu^-\ ,\ \mu^- \pi^+\ .
	\]
	The decay width of these channels can be found in~\cite{Atre09}.	

	It is possible to evaluate the flux of the sterile neutrinos within reasonable approximation. 
	The flux of light neutrinos $\nu_\alpha$, like the one presented in Fig.~\ref{fig:branch}, %
	produced from the leptonic or semi-leptonic decay of a given parent particle %
	$P \ra \nu_\alpha + X$ can be suitably scaled to obtain a valid heavy neutrino flux.
	Neglecting energy dependence, a reliable scale factor is given by the ratio with respect to the SM %
	of the partial decay widths in the model with an extra neutrino.
	The ratio will be proportional to the mixing parameter and will contain only kinematic %
	functions of the Majorana neutrino mass, $m_N$.
	With this in mind, the corresponding flux for the sterile neutrino is estimated as follows:
	\begin{equation}
		\label{eq:scaleflux}
		\dv{\phi_N}{E} (E_N) \approx \sum_{\alpha, P}  \mathcal{K}^P_\alpha(m_N)\ \dv{\phi_{P\ra\nu_\alpha}}{E} (E_N - m_N)
		\quad , \quad 
		\mathcal{K}^P_{\alpha}(m_N) \equiv \frac{\Gamma_{\text{SM}+N}}{\Gamma_\text{SM}}\ .
	\end{equation}
	The scale factors $\mathcal{K}$ can be computed analytically for leptonic two-body decays of pseudo-scalar mesons~\cite{Shrock81}.
	This ratio in the case of two-body decays is responsible for correcting phase space and helicity terms.
	As a matter of fact, a Majorana neutrino is not affected by helicity suppression, and as a consequence the %
	electronic channels are enhanced with respect to the muonic.
	The factors are shown on the right plot of figure Fig.~\ref{fig:branch}.
	In the case of three-body decays the computation is numerical, and the factors $\mathcal{K}$ do not show any enhancement %
	but account just for the correct phase space.

% Head 3
\section{The Near Detector}

	\begin{table}
		\small
		\centering
		\begin{tabular}{ccrr}
			\toprule
				& Channel	& BR (\%)	& $m_N$(MeV) \\
			\hline
			$\pi^+$	& $\mu^+ \nu_\mu$	& \np{99.98}		& \np{33.91}	\\
				& $e^+ \nu_e$		& \np{0.01}		& \np{139.06}	\\
			\hline
			$K^+$	& $\mu^+ \nu_\mu$	& \np{63.56}		& \np{387.81}	\\
				& $\pi^0 e^+ \nu_e$	& \np{5.07}		& \np{358.19}	\\
				& $\pi^0 \mu^+ \nu_\mu$	& \np{3.35}		& \np{253.04}	\\
				& $e^+ \nu_e$		& \np{0.16}		& \np{493.17}	\\
			\hline
			$K^0_L$	& $\pi^\pm e^\mp\nu_e$		& \np{40.55}	& \np{357.12}	\\
				& $\pi^\pm\mu^\mp\nu_\mu$	& \np{27.04}	& \np{252.38}	\\ 
			\hline
			$\mu^+$	& $\cj{\nu_\mu}e^+ \nu_e$	&\np{100.00}	& \np{105.14}	\\
			\bottomrule
		\end{tabular}
		\hspace{2em}
		\begin{tabular}{l@{\ }c@{\ }c@{\ }c@{\ }}
			\toprule
				& \textbf{PS191}	& \textbf{LArTPC}	& \textbf{HPArFGT}		\\
			\midrule
			Baseline& 128~m			& 574~m		& 578~m			\\
			Sizes	& --	& 3m$\times$3m$\times$4m & 3.5m$\times$3.5m$\times$6.4m	\\
			Volume	& 216~m${}^3$	& 36~m${}^3$	& 78.4~m${}^3$	\\
			Weight	& -- 		& 50~ton	& 8~ton		\\
			POT	& \np{0.86e19}	& \np{13.23e21}	& \np{13.23e21}	\\
			Exposure& 1.0 		& 12.7		& 27.4	\\
			\bottomrule
		\end{tabular}
		\caption{Left: Main decay channels yielding neutrinos in a fixed target experiment.
			Only secondary particles up to the neutral kaon are shown.
			Right: Features of the two detectors which will form the ND system, compared to PS191.
			The exposure is defined as POT$\times$Volume$\times$Baseline${}^{-2}$ with respect to PS191.
			For this search, volume is the driving feature, whereas the fiducial weight affects the background.}
		\label{tab:nd}
	\end{table}

	The upcoming DUNE experiment will study oscillation of neutrinos in great detail~\cite{DUNE}, thanks to %
	the 40~kton Liquid Argon Time Projection Chamber (LArTPC) situated \np{1300}~km from the proton target.
	A Near Detector (ND) is nevertheless needed in order to normalise the flux of neutrinos reaching the far detector: %
	the ND will be placed \np{574}~m from the target.
	Even if the final design of the ND has not been decided yet, its active volume will be exposed to an extremely intense %
	neutrino beam during data runs, thanks to its proximity to the target and its size.
	The ND will likely be a hybrid concept, consisting of a LArTPC placed in front of a High Pressure Argon TPC (HPArTPC).
	A summary of the features of the two detector is reported in Tab.~\ref{tab:nd}, where they are compared to PS191.

	The expected number of heavy neutrino decays inside the detector for a given channel $d$ can be naively evaluated using the following formula:

	\vspace{0.5em}
	\noindent \begin{minipage}{0.5\linewidth}
		\begin{equation}
			\label{eq:event}
			\mathcal{N}_d = \int \dd{E}\ P_d(E) W_d(E)\, \dv{\phi_N}{E} (E) \ ,
		\end{equation}
	\end{minipage}
	\begin{minipage}{0.5\linewidth}
		\begin{equation}
			P_d(E) = e^{-\frac{\Gamma_\text{tot} L}{\gamma \beta}} %
			\qty(1-e^{-\frac{\Gamma_\text{tot} \lambda}{\gamma \beta}}) \frac{\Gamma_d}{\Gamma_\text{tot}}\ , 
		\end{equation}
	\end{minipage}
	\vspace{0.1em}

	\noindent where $W_d$ is an efficiency factor derived from background reduction analyses, and $\dv*{\phi_N}{E}$ %
	is the number of $N$ expected at the ND.
	The term $P_d$ accounts for the probability of a Majorana neutrino of energy $E$ to travel the baseline %
	distance and decay inside the near detector, where $L$ denotes the baseline, $\lambda$ the length of detector, %
	$\Gamma_d$ the partial decay width for the mode $d$ and $\Gamma_\text{tot}$ the total decay width.

	Except from $N$ decaying into three neutrinos, all the other decay channels are detectable.
	Ordinary neutrino-nucleon interaction occurring within the fiducial volume of the detector are the major source of noise.
	The most copious events are quasi-elastic interactions, either via charged currents (CCQE) or via neutral currents (NCE), %
	deep inelastic scattering (DIS) and resonance pion production (CC1$\pi$).
	In order to reduce the background, a million neutrino-nucleon events are generated with \textsc{genie}~\cite{genie} %
	and passed to a fast Monte Carlo which accounts for detector smearing effects and mis-identification.
	Events with detected hadronic activity are discarded, for it is a conclusive evidence for a beam-related scattering event.
	The kinematic properties of the remaining background events are compared to simulated signal events: %
	the different behaviours of the two further help in discriminating signal events from the background.
	Conservative data analysis cuts are applied to maximise the background reduction, while retaining as many signal events as possible.
	The weighting factors $W_d(E)$ in Eq.~\ref{eq:event} are the binned ratio of the $N$ energy spectrum after and before the analysis cuts.	
	
% Head 4
\section{Results}

	\begin{figure}
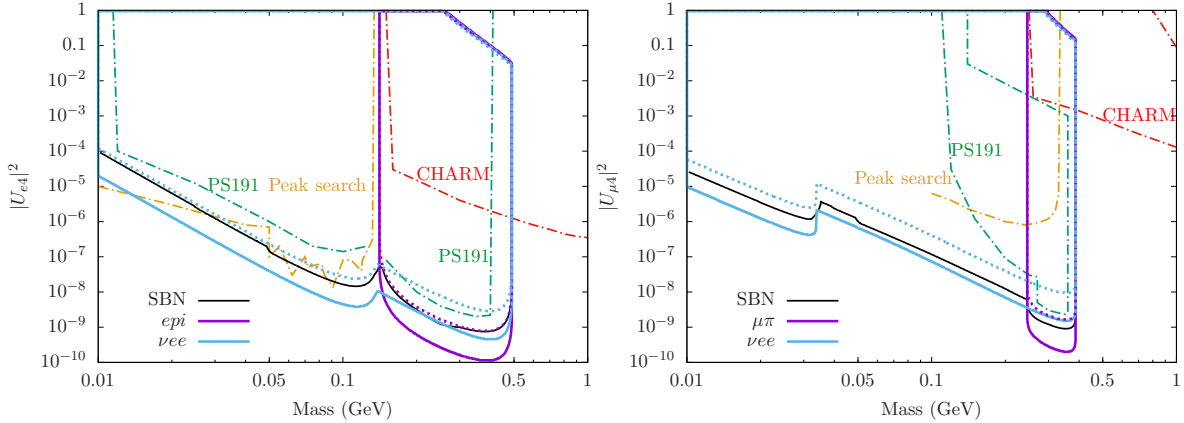

		\centering
		\resizebox{0.5\linewidth}{!}{\input{sens_e_simple.tex}}
		\hspace{-2em}
		\resizebox{0.5\linewidth}{!}{\input{sens_m_simple.tex}}
		\footnotesize
		\caption{Sensitivity of the ND system to direct search of right-handed neutrinos, 
			studying the decay modes $N \ra e \pi$ and $N \ra \nu e e$ for $|U_{e 4}|^2$ (left) %
			and the decay modes $N \ra \mu \pi$ and $N \ra \nu e e$ for $|U_{\mu 4}|^2$ (right).
			The solid lines corresponds to the analysis before the background analysis ($W_d = 1$).
			The dotted lines are drawn after the background analysis.
			The current limits from PS191~\cite{Bernardi88}, CHARM~\cite{Charm} and peak searches~\cite{Peak} are also shown for comparison.}
		\label{fig:sens}
	\end{figure}

	The plots in Fig.~\ref{fig:sens} are the combined 90\,\% C.L.\ exclusion lines for the ND system (LArTPC + HPArTPC), %
	defined following the procedure described in~\cite{Feldman97}.
	For both mixings $|U_{e 4}|^2$ and $|U_{\mu 4}|^2$, the plots show the results for the channels with good detection prospects, %
	i.e.\ the modes $e^- \pi^+$ for $|U_{e 4}|^2$, $\mu^- \pi^+$ for $|U_{e 4}|^2$, and $\nu e^+ e^-$ for both.
	The solid lines corresponds to the background-less analysis ($W_d = 1$), while the dashed lines are the sensitivities after %
	the background reduction.
	The prediction assuming vanishing background for the future Short Baseline (SBN) program performed in a similar study~\cite{Ballett17} is also overlaid.
	In the absence of signal, the ND will be able to improve the existing limits for both mixing parameters.
	An overall improvement up to one order of magnitude in the sensitivity is expected for all the mass range with respect to %
	previous and near-future experiments.
	The other decay channels have been studied as well, leading to similar conclusions.

% References

\end{document}